\documentclass[12pt]{iopart}
\usepackage{amssymb,graphicx,cite}
\begin{document}
\title[Sextic potential for $\gamma$-rigid prolate nuclei]{Sextic potential for $\gamma$-rigid prolate nuclei}
\author{P. Buganu$^{1}$ and R. Budaca$^{1}$}
\address{$^{1}$Department of Theoretical Physics, Horia Hulubei National Institute of Physics and Nuclear Engineering, Str. Reactorului 30, RO-077125, POB-MG6, Bucharest-M\v{a}gurele, Romania}
\eads{\mailto{buganu@theory.nipne.ro}, \mailto{rbudaca@theory.nipne.ro}}
\begin{abstract}

The equation of the Bohr-Mottelson Hamiltonian with a sextic oscillator potential is
solved for $\gamma$-rigid prolate nuclei. The associated shape phase space is reduced to three
variables which are exactly separated. The angular equation has the spherical harmonic
functions as solutions, while the $\beta$ equation is brought to the quasi-exactly solvable
case of the sextic oscillator potential with a centrifugal barrier. The energies and the
corresponding wave functions are given in closed form and depend, up to a scaling factor, on a single parameter. The $0^{+}$ and $2^{+}$ states are exactly determined, having an important role in the assignment of some ambiguous states for the experimental $\beta$ bands. Due to the special properties of the sextic potential, the model can simulate, by varying the free parameter, a shape phase transition from a harmonic
to an anharmonic prolate $\beta$-soft rotor crossing through a critical point.  Numerical
applications are performed for 39 nuclei:  $^{98-108}$Ru, $^{100,102}$Mo, $^{116-130}$Xe, $^{132,134}$Ce, $^{146-150}$Nd, $^{150,152}$Sm, $^{152,154}$Gd, $^{154,156}$Dy, $^{172}$Os, $^{180-196}$Pt, $^{190}$Hg and $^{222}$Ra. The best candidates for the critical point are found to be $^{104}$Ru and
$^{120,126}$Xe, followed closely by $^{128}$Xe, $^{172}$Os, $^{196}$Pt and $^{148}$Nd.

\end{abstract}


\vspace{2pc}

\submitto{\jpg}

\section{Introduction}

Even if there are more than 60 years since the Bohr-Mottelson model \cite{Bohr1,Bohr2} was proposed, the interest in its solutions has grown exponentially in the last two decades and this happened at least from two main reasons. One is represented by the appearance of its cornerstone solutions, called E(5) \cite{Iachello1}, X(5) \cite{Iachello2}, Y(5) \cite{Iachello3} and Z(5) \cite{Bonatsos1}, which describe the nuclei situated in the critical points of the shape phase transitions from spherical vibrator to a $\gamma$-unstable system, from spherical vibrator to prolate rotor, from axial rotor to triaxial rotor, and from prolate rotor to oblate rotor, respectively. The second reason is due to the increasing number of the experimental data for the quadrupole low-lying states and which require interpretation. The multitude of various solutions for the Bohr-Mottelson Hamiltonian proposed for this scope were systematically reviewed in Refs. \cite{Fortunato1,Casten,Casten2,Prochniak,Cejnar,Rowe,Isacker}.

Soon after the critical point symmetries were proposed, another direction started to develop by checking if the $\gamma$-rigid versions of some of these symmetries reveal new features for the field of the nuclear shape phase transitions. Therefore, the Z(4) \cite{Bonatsos2} and X(3) \cite{Bonatsos3} solutions were indicated as $\gamma$-rigid counterparts of the Z(5) and X(5) models, the first corresponding to a triaxial shape and the second to a prolate shape. Z(4) is described by the Davydov-Chaban Hamiltonian \cite{Davydov} which emerges from the Bohr-Mottelson Hamiltonian by freezing the $\gamma$ degree of freedom. Going further and imposing axial symmetry, one obtains after quantization in the remaining curvilinear coordinates a new Hamiltonian \cite{Bonatsos3} associated to the X(3) equation. Both Z(4) and X(3) involve an infinite square well potential (ISWP) for the $\beta$ variable, a potential which is well known in literature to be a fair approximation for the critical point potential of the spherical to deformed shape phase transition. Between the advantages of the ISWP are included the simple and the parameter free analytical solution. However, these approaches lack in flexibility and consequently in experimental candidates. A question that naturally arises is if there are other more flexible potentials which can closely reproduce the properties of the ISWP and in the same time present new features, but still lead  to analytical solutions and depend on very few parameters. The quartic anharmonic oscillator potential (QAOP) \cite{Budaca1} and the sextic anharmonic oscillator potential (SAOP) \cite{Budaca2} applied for X(3) and the quasi-exactly solvable sextic potential \cite{Buganu1} used in connection to Z(4) are good examples of this challenge's importance.

In the present work the ISWP used in Ref. \cite{Bonatsos3} for X(3) is replaced by the sextic oscillator potential \cite{Ushveridze}. It is worth to mention here that firstly the Bohr-Mottelson Hamiltonian with sextic potential was solved for $\gamma$-unstable nuclei \cite{Levai1,Levai2}, then for $\gamma$-stable triaxial nuclei \cite{Raduta1,Buganu2} and ultimately  for $\gamma$-stable prolate nuclei \cite{Raduta2}. Other numerical applications of these solutions were done in Refs. \cite{Kharb,Raduta3,Buganu3}. In the present case, the variables are exactly separated and the resulted $\beta$ equation is quasi-exactly solved. The energy spectra and the wave functions are given in an analytical form depending, up to a scale factor, on a single free parameter. For particular values of the free parameter, some terms of the sextic potential as $\beta^{2}$ or $\beta^{4}$ vanish leading to parameter free solutions. Due to the special properties of the sextic potential, by varying the free parameter, a shape phase transition can be covered from a $\gamma$-rigid prolate harmonic vibrator to an anharmonic one, passing through a critical point. The energies and the $B(E2)$ transition rates associated to the states $0^{+}$ and $2^{+}$ of the ground band and of the first two $\beta$ bands are exactly determined, representing thus reliable reference signatures for the identification of the possible $\beta$ bands. For the rest of the states an approximation is involved whose consistency is validated in respect to the numerical results of the previous solutions of the same shape phase space, {\it i.e.} X(3), X(3)-$\beta^{2}$ \cite{Budaca1}, X(3)-$\beta^{4}$ \cite{Budaca1} and X(3)-$\beta^{6}$ \cite{Budaca2}. The solution proposed in the present work is conventionally called X(3)-Sextic, in connection to the X(3) model. Numerical results of  X(3)-Sextic, for the energy spectra of the ground band and of the first two $\beta$ bands as well as for several  $E2$ transition probabilities, are compared with the available experimental data for the nuclei: $^{98-108}$Ru, $^{100,102}$Mo, $^{116-130}$Xe, $^{132,134}$Ce, $^{146-150}$Nd, $^{150,152}$Sm, $^{152,154}$Gd, $^{154,156}$Dy, $^{172}$Os, $^{180-196}$Pt, $^{190}$Hg and $^{222}$Ra. A special attention is allocated to the experimental evidence of the shape phase transition from $\gamma$-rigid prolate harmonic vibrator to an anharmonic one and to the identification of the heads of the $\beta$ bands, respectively.

The plan of the present work is the following: X(3)-Sextic solution is described in detail in Section II, while in Section III its numerical applications are presented and the results are discussed from both theoretical and experimental point of view. Finally, the main results of X(3)-Sextic are collected in Section IV. For a better understanding and future applications of the X(3)-Sextic solution,  an appendix with explicit expressions for energies and wave functions is  added.

\renewcommand{\theequation}{2.\arabic{equation}}
\section{X(3)-Sextic solution}
\label{sec:2}
\subsection{Separation of variables}

The Bohr-Mottelson model \cite{Bohr1,Bohr2} with $\gamma$ rigidity ($\dot{\gamma}=0$) leads to the Davydov-Chaban Hamiltonian \cite{Davydov}, while in a more particular situation of the $\gamma$-rigid prolate shape ($\gamma=0^{\circ}$) it takes the following form \cite{Bonatsos3,Sitenko,Eisenberg}:
\begin{equation}
H=-\frac{\hbar^{2}}{2B}\left[\frac{1}{\beta^{2}}\frac{\partial}{\partial\beta}\beta^{2}\frac{\partial}{\partial\beta}+\frac{1}{3\beta^{2}}\Delta_{\theta,\phi}\right]+V(\beta),
\end{equation}
where,
\begin{equation}
\Delta_{\theta,\phi}=\frac{1}{\sin\theta}\frac{\partial}{\partial\theta}\sin\theta\frac{\partial}{\partial\theta}+\frac{1}{\sin^{2}\theta}\frac{\partial^{2}}{\partial\phi^{2}}.
\end{equation}
The separation of variables is achieved choosing the wave function as $\Psi(\beta,\theta,\phi)=F(\beta)Y_{L,M_{L}}(\theta,\phi)$:
\begin{equation}
-\Delta_{\theta,\phi}Y_{L,M_{L}}(\theta,\phi)=L(L+1)Y_{L,M_{L}}(\theta,\phi),
\end{equation}
\begin{equation}
\left[-\frac{1}{\beta^{2}}\frac{d}{d\beta}\beta^{2}\frac{d}{d\beta}+\frac{L(L+1)}{3\beta^{2}}+v(\beta)\right]F(\beta)=\varepsilon F(\beta),
\label{eqbeta}
\end{equation}
where by $v(\beta)=2BV(\beta)/\hbar^{2}$ and $\varepsilon=2BE/\hbar^{2}$ are denoted the reduced potential and energy. Eq. (\ref{eqbeta}) was solved before for an ISWP \cite{Bonatsos3}, a pure harmonic oscillator (HO) \cite{Budaca1}, a QAOP \cite{Budaca1} and a SAOP \cite{Budaca2}. In the following, a sextic potential will be considered \cite{Ushveridze}.  As can be seen from Table \ref{tab} where a comprehensive account of $\gamma$-rigid solutions is presented, the sextic potential contains both $\beta^{6}$ and $\beta^{4}$ terms compared to QAOP and SAOP. On the other hand, all these three potentials, for particular values of the corresponding free parameters, reproduce closely the ISWP and the HO.
\begin{table}
\caption{The potentials in the $\beta$ variable and the $\gamma$ rigidity values for the relevant $\gamma$-rigid solutions.}
\begin{center}
\begin{tabular}{llc}
\hline
\hline
Solution&$\beta$ potential&$\gamma$\\
\hline
X(3)&0,$\;$if$\;$$\beta\leq\beta_{\omega}$,$\;$&$0^{\circ}$\\
    &$\infty$,$\;$if$\;$$\beta>\beta_{\omega}$&\\
X(3)-$\beta^{2}$&$\sim\beta^{2}$&$0^{\circ}$\\
X(3)-$\beta^{4}$&$\sim\beta^{4}$&$0^{\circ}$\\
X(3)-$\beta^{6}$&$\sim\beta^{6}$&$0^{\circ}$\\
QAOP&$\frac{1}{2}\alpha_{1}\beta^{2}+\alpha_{2}\beta^{4},$&$0^{\circ}$\\
&$\alpha_{1}\geq0,\;\alpha_{2}>0$&\\
SAOP&$\frac{1}{2}\alpha_{1}\beta^{2}+\alpha_{2}\beta^{6},$&$0^{\circ}$\\
&$\alpha_{1}\geq0,\;\alpha_{2}>0$&\\
X(3)-Sextic&$(b^{2}-4ac)\beta^{2}+2ab\beta^{4}+a^{2}\beta^{6}$,&$0^{\circ}$\\
&$c,a>0,\;b\in R$&\\
Z(4)&0,$\;$if$\;$$\beta\leq\beta_{\omega},$&$30^{\circ}$\\
&$\infty$,$\;$if$\;$$\beta>\beta_{\omega}$&\\
Z(4)-$\beta^{2}$&$\sim\beta^{2}$&$30^{\circ}$\\
Z(4)-Sextic&$(b^{2}-4ac)\beta^{2}+2ab\beta^{4}+a^{2}\beta^{6}$,&$30^{\circ}$\\
&$c,a>0,\;b\in R$&\\
\hline
\hline
\end{tabular}
\end{center}
\label{tab}
\end{table}

\subsection{Solution of the $\beta$ equation with sextic potential}

\begin{figure}
\begin{center}
\includegraphics[width=1\textwidth]{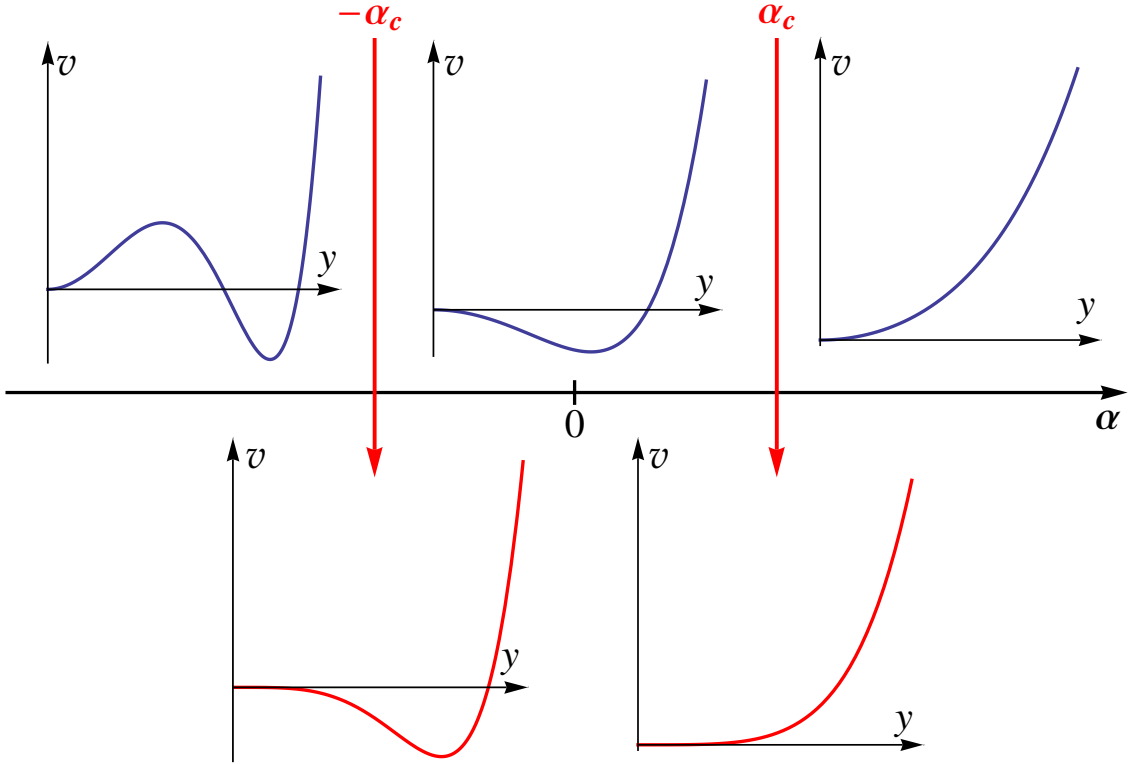}
\end{center}
\caption{The shape evolution of the scaled sextic potential $v_{0}^{(2)}(y)$, given by Eq. (\ref{poty}), as a function of the free parameter $\alpha$ for $c_0^{(2)}=13/4$ and $u_{0}^{(2)}=0$.}
\label{fig1}
\end{figure}

In what follows, it is preferable to write Eq. (\ref{eqbeta}) in a Schr\"{o}dinger form. This is realized by changing the wave function as $F(\beta)=\beta^{-1}\varphi(\beta)$:
\begin{equation}
\left[-\frac{d^{2}}{d\beta^{2}}+\frac{L(L+1)}{3\beta^{2}}+v(\beta)\right]\varphi(\beta)=\varepsilon\varphi(\beta).
\label{eqsch}
\end{equation}
Further, Eq. (\ref{eqsch}) is brought to a form associated with the quasi-exactly case of the sextic oscillator potential with a centrifugal barrier \cite{Ushveridze} by considering the relations:
\begin{equation}
\frac{L(L+1)}{3}=\left(2s-\frac{1}{2}\right)\left(2s-\frac{3}{2}\right),
\label{eqsl}
\end{equation}
\begin{equation}
v(\beta)=\left[b^{2}-4a\left(s+\frac{1}{2}+M\right)\right]\beta^{2}+2ab\beta^{4}+a^{2}\beta^{6},\;M=0,1,2,...
\label{vbeta}
\end{equation}
Here by \textit{quasi-exactly solvable} one  understands that the eigenvalue problem is exactly solved only for a part of the spectrum, namely for $M+1$ eigenstates.
The sextic potential (\ref{vbeta}) depends on two parameters, $a$ and $b$, and on an integer $M$, but also on  the quantum number $L$ through $s$. Apparently, the potential is state dependent, but, as it will be shown, $M$ can be used instead to keep the potential independent on $L$ by forcing that:
\begin{equation}
s+\frac{1}{2}+M=const.\equiv c.
\label{const}
\end{equation}
Expressing $s$ from Eq. (\ref{eqsl}) as a function of $L$,
\begin{equation}
s(L)=\frac{1}{2}\left[1+\sqrt{\frac{L(L+1)}{3}+\frac{1}{4}}\right],
\label{sl}
\end{equation}
it is observed that the condition (\ref{const}) is impossible to be fulfilled  for $L\geq4$, $s$ taking irrational values while $M$ is an integer number. Instead, for $L=0$ and $2$ the expression of $s$ given by Eq. (\ref{sl}) reduces to a simpler form:
\begin{equation}
s'(L)=\frac{L+3}{4}.
\label{sl02}
\end{equation}
By comparing, in Table \ref{tab1}, $s(L)$ and $s'(L)$ for different values of $L$ it can be appreciated that the latter represents a good approximation of the former at least for $L\leq10$.
\setlength{\tabcolsep}{8.5pt}
\begin{table}[ht!]
\caption{A comparison of $s(L)$ and $s'(L)$, given by Eqs. (\ref{sl}) and (\ref{sl02}) for even $L\leq10$.}
\vspace{0.3cm}
\begin{center}
\begin{tabular}{ccccccc}
\hline
\hline
$L$&0&2&4&6&8&10\\
\hline
$s$&0.75&1.25&1.81&2.39&2.96&3.54\\
$s'$&0.75&1.25&1.75&2.25&2.75&3.25\\
\hline
\hline
\end{tabular}
\end{center}
\label{tab1}
\end{table}
Using $s'(L)$ even for $L\geq4$, the condition (\ref{const}) becomes
\begin{equation}
c=M+\frac{L}{4}+\frac{5}{4},
\end{equation}
which is constant if $M$ decreases by one unit while $L$ increases by four units:
\begin{eqnarray}
(M,L)&:&(K,0),(K-1,4),(K-2,8),...\Rightarrow K+\frac{5}{4}=c^{(K)}_0,\\
(M,L)&:&(K,2),(K-1,6),(K-2,10),...\Rightarrow K+\frac{7}{4}=c^{(K)}_2.
\label{rule1}
\end{eqnarray}
Here by $K$ is denoted the maximum value of $M$. Finally remain two potentials independent on $L$, one for $L=0,4,8,...$ and other for $L=2,6,10,...$, which slightly differ through the coefficient of $\beta^{2}$ by the quantity $c^{(K)}_2-c^{(K)}_0=1/2$. The sextic potential equation (\ref{eqsch}) can be more simplified by reducing the number of  parameters through the change of variable $\beta=ya^{-1/4}$ and adopting the notations $\alpha=b/\sqrt{a}$ and $\varepsilon_{y}=\varepsilon/\sqrt{a}$:
\begin{equation}
\left[-\frac{d^{2}}{dy^{2}}+\frac{L(L+1)}{3y^{2}}+v_{m}^{(K)}(y)\right]\eta(y)=\varepsilon_{y}\eta(y),
\label{eqy}
\end{equation}
where,
\begin{equation}
v_{m}^{(K)}(y)=(\alpha^{2}-4c^{(K)}_m)y^{2}+2\alpha y^{4}+y^{6}+u^{(K)}_m, \,m=0,2.
\label{poty}
\end{equation}
Here, $u^{(K)}_m$ are constants added in order to minimize the slight difference between the two potentials (\ref{poty}). Thus, $u^{(K)}_m$ are fixed such that the potentials to have the same minimum energy:
\begin{eqnarray}
u_{0}^{(K)}=0,\;u_{2}^{(K)}&=&\left(\alpha^{2}-4c_0^{(K)}\right)\left(y_{0,0}^{(K)}\right)^{2}-\left(\alpha^{2}-4c_2^{(K)}\right)\left(y_{0,2}^{(K)}\right)^{2}\nonumber\\
&&+2\alpha\left[\left(y_{0,0}^{(K)}\right)^{4}-\left(y_{0,2}^{(K)}\right)^{4}\right]+\left(y_{0,0}^{(K)}\right)^{6}-\left(y_{0,2}^{(K)}\right)^{6},
\end{eqnarray}
where by $y_{0,m}^{(K)}$ are denoted the extreme points of the scaled potential (\ref{poty}):
\begin{equation}
(y_{0,m}^{(K)})^{2}=\frac{1}{3}\left(-2\alpha\pm\sqrt{\alpha^{2}+12c_m^{(K)}}\right),\,m=0,2.
\end{equation}
An elegant representation of the shape evolution of the sextic potential (\ref{poty}) as a function of the free parameter $\alpha$ is shown in Fig. \ref{fig1}.
Thus, coming from $\alpha\rightarrow\infty$ towards $\alpha=0$ through the point $\alpha_c$, a shape phase transition is covered from a harmonic vibrator to an anharmonic one crossing a critical point for which the potential is flat. Going farther to $\alpha\rightarrow-\infty$ a simultaneous spherical and deformed minima appear separated by a maximum. Actually this representation is a projection of the sextic potential evolution from the plane of parameters $(a,b)$ \cite{Levai2} on the real axis of the free parameter $\alpha$. Indeed, in Ref. \cite{Levai2} the critical point corresponds to the parabola $a=b^{2}/4c$, while here is given by $\alpha^{2}=4c$, where $\alpha=b/\sqrt{a}$. Another important remark is that points $\alpha_c=\pm2\sqrt{c}$ are moving away from origin as the $K$ value increases.

So far, all the above efforts were done to bring Eq. (\ref{eqsch}) with a sextic potential \cite{Ushveridze} to a solvable equation for the $\gamma$-rigid prolate nuclei and then to have a more convenient form of it by reducing the number of parameters. Further, the attention will be focused on the solution of Eq. (\ref{eqy}). Therefore, choosing as an ansatz the function \cite{Ushveridze}
\begin{equation}
\eta^{(M)}(y)\sim P^{(M)}(y^2)y^{2s'-\frac{1}{2}}e^{-\frac{y^4}{4}-\frac{\alpha y^{2}}{2}},
\label{ansatz}
\end{equation}
Eq. (\ref{eqy}) with the potential (\ref{poty}) is reduced to the equation \cite{Ushveridze}
\begin{eqnarray}
\left[-\left(\frac{d^{2}}{dy^{2}}+\frac{4s'-1}{y}\frac{d}{dy}\right)+2\alpha y\frac{d}{dy}+2y^{2}\left(y\frac{d}{dy}-2M\right)\right]P^{(M)}(y^{2})&=&\nonumber\\
=\lambda P^{(M)}(y^{2}),&&
\label{eqQ}
\end{eqnarray}
where $P^{(M)}$ are polynomials in $y^{2}$ of order $M$.  Writing Eq. (\ref{eqQ}) in a matrix form as in the Appendix of Ref. \cite{Raduta1}, the eigenvalues $\lambda$ are obtained as
\begin{equation}
\lambda\equiv\lambda_{n,L}^{(K)}=\varepsilon_{y}-4\alpha s'-u_{m}^{(K)}-\frac{1}{\langle y^{2}\rangle}\frac{L}{6}\left(\frac{L}{2}-1\right),
\label{enl}
\end{equation}
where the last term, containing the mean value of $y^2$, was extracted from the centrifugal term of Eq. (\ref{eqy}) such that $s(L)$ could be approximated by $s'(L)$ for $L\geq4$. By adding this term to the energies (\ref{enl}), arises the possibility to describe very well also the states with $L^{+}>10^{+}$, for which $s(L)$ and $s'(L)$ otherwise  start to diverge.
Finally, the total energy of the system is easily deduced from Eq. (\ref{enl}):
\begin{equation}
E_{n,L}=\frac{\hbar^{2}\sqrt{a}}{2B}\left[\lambda_{n,L}^{(K)}(\alpha)+\alpha(L+3)+u_{m}^{(K)}(\alpha)+\frac{1}{\langle                                                                                                                                                                                                                                                                                                                                                                                                                                                          y^{2}\rangle}\frac{L}{6}\left(\frac{L}{2}-1\right)\right].
\label{ener}
\end{equation}
The total energy (\ref{ener}) depends on two quantum numbers, the order of the wave function's zero $n$ and the total angular momentum $L$, on a fixed integer $K$, on a scale factor $(\hbar^{2}\sqrt{a})/(2B)$ and on a free parameter $\alpha$. There are two possibilities for numerical applications to the experimental data. The first one is to normalize the energies (\ref{ener}) to the ground state energy and to fit both the scale factor and the free parameter or the second one by normalizing to the first excited state energy, as
\begin{equation}
R_{n,L}^{(K)}=\frac{E_{n,L}-E_{0,0}}{E_{0,2}-E_{0,0}},
\label{enrat}
\end{equation}
and then to fit only the free parameter. In the present paper, the second method is adopted with $n=0$, $n=1$ and $n=2$ corresponding to the ground band, the first and second $\beta$ bands, respectively.

\subsection{$E2$ transition probabilities}
Another important aspect of the present approach comparing to the previous ones, {\it i.e.} QAOP \cite{Budaca1} and (SAOP) \cite{Budaca2} is that its associated quasi-exactly solvable method allows to determine the expressions for the wave functions. In this way, numerical results for $B(E2)$ transition rates can be performed. The total wave function is a product of the radial and angular components:
\begin{equation}
\Psi_{n,L,M_{L}}^{(M)}(\beta,\theta,\phi)=F_{n,L}^{(M)}(\beta)Y_{L,M_{L}}(\theta,\phi),
\end{equation}
where $Y_{L,M_{L}}(\theta,\phi)$ are the well known spherical harmonics, while $F_{n,L}^{(M)}(\beta)$ are expressed in terms of the normalized ansatz functions (\ref{ansatz}). The harmonic transition operator for $\gamma$-rigid prolate case gets the simplified form \cite{Bonatsos3}:
\begin{equation}
T_{\mu}^{(E2)}=t\beta\sqrt{\frac{4\pi}{5}}Y_{2,\mu}(\theta,\phi),
\label{tranop}
\end{equation}
where $t$ is a scale parameter which is dropped for normalized $B(E2)$ values. Therefore, the final expression of the reduced $E2$ transition probabilities, normalized to the transition from the first excited state to the ground state, is:
\begin{equation}
T_{n,L,n',L'}=\frac{B(E2;n,L\rightarrow n',L')}{B(E2;0,2\rightarrow0,0)}=\left(\frac{C_{\;000}^{L2L'}I_{n,L;n',L'}}{C_{\;000}^{220}I_{0,2;0,0}}\right)^{2},
\label{tranrap}
\end{equation}
where the radial matrix element $I_{n,L;n',L'}$ can be given either in $\beta$ or $y$ variable:
\begin{eqnarray}
I_{n,L;n',L'}&=&\int_{0}^{\infty}F_{n,L}^{(M)}(\beta)\beta F_{n',L'}^{(M)}(\beta)\beta^{2}d\beta\nonumber\\
&=&\int_{0}^{\infty}\varphi_{n,L}^{(M)}(\beta)\beta\varphi_{n',L'}^{(M)}(\beta)d\beta\\
&=&a^{-1/4}\int_{0}^{\infty}\eta_{n,L}^{(M)}(y)y\eta_{n',L'}^{(M)}(y)dy.\nonumber
\end{eqnarray}

As it will be shown further, X(3)-Sextic presents unique properties from both theoretical and experimental point of view by comparison with the previous similar solutions, making it a powerful tool in the description of the $\gamma$-rigid prolate nuclei.

\renewcommand{\theequation}{3.\arabic{equation}}
\section{Numerical results and discussion }
\label{sec:3}

\subsection{Theoretical aspects of X(3)-Sextic}

\begin{figure}
\begin{center}
\includegraphics[width=0.75\textwidth]{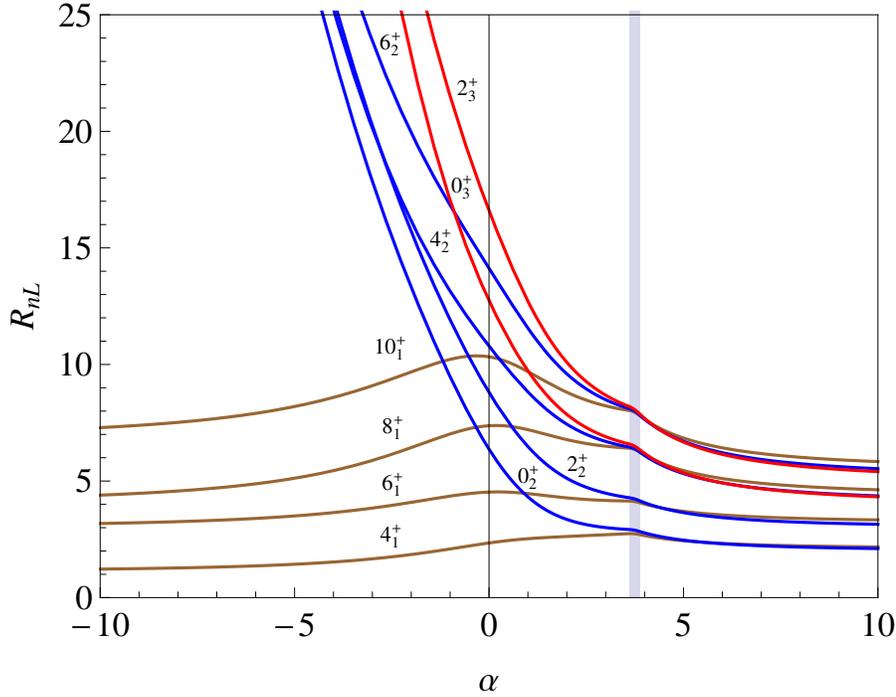}
\end{center}
\caption{The energy spectra of the ground band and of the first two $\beta$ bands, given by Eq. (\ref{enrat}), are plotted as a function of the free parameter $\alpha$ in the interval $\alpha\in[-10,10]$ for $K=2$. The energy lines are indexed by $L_{n+1}^{+}$, where $n=0$ for the ground band, $n=1$ for the first $\beta$ band and $n=2$ for the second $\beta$ band.}
\label{fig2}
\end{figure}

In the following, the entire discussion will be restricted to the case $K=2$ for which, as can be deduced from Eq. (\ref{rule1}), the properties of the most
representative low-lying states of the ground band (up to $10^{+}$), the first $\beta$ band (up to $6^{+}$)  and the second $\beta$ band (up to $2^{+}$) are reproduced. At this point, the index $K$ will be dropped as it has a fixed value. If necessary, results for $K>2$ can be considered without any problems \cite{Raduta1}. The beauty of the case $K=2$ is that explicit analytical expressions for energies and wave functions are still possible. For a better understanding and possible future applications of X(3)-Sextic solution, the cases corresponding to $M=0$, $M=1$ and $M=2$ are detailed in the Appendix. A plot of the energies given by Eq. (\ref{enrat}) for $K=2$ and $\alpha\in[-10,10]$ are presented in Fig. \ref{fig2} from which one can draw some useful insights. For $\alpha\rightarrow\infty$, X(3)-Sextic reproduces the results of the $\gamma$-rigid prolate harmonic vibrator called X(3)-$\beta^{2}$ \cite{Budaca1}, while for $\alpha$ close to zero a $\gamma$-rigid prolate anharmonic vibrator shows up. Special situations appear for $\alpha\rightarrow -\infty$, where the ground band energies maintain finite values while the $\beta$ band energies go to infinity, and for $\alpha\in[2\sqrt{c_{0}},2\sqrt{c_{2}}]$ (the narrow vertical area) where a critical point of a first order shape phase transition occurs. In the region $\alpha\in[-2\sqrt{c_{0}},-\infty]$, as can be seen from Fig. \ref{fig1}, simultaneous spherical and deformed minima appear separated by a high barrier. On the other hand, in the interval $\alpha\in[2\sqrt{c_{0}},2\sqrt{c_{2}}]$ the potential is completely flat as an infinite square well potential, which is known from the literature that usually corresponds to critical points. Indeed, by analyzing the first derivative of the energy (\ref{enrat}) in respect to the free parameter $\alpha$ it is found that it has a discontinuity for an $\alpha_c\in[2\sqrt{c_{0}},2\sqrt{c_{2}}]$. The interpretation of this result is that a first order shape phase transition takes place between  a $\gamma$-rigid prolate harmonic vibrator and a $\gamma$-rigid prolate anharmonic vibrator. Actually this is not a surprise, similar results for different shape phase transitions being reported in Refs.  \cite{Buganu1,Levai2,Yigitoglu,Casten,Turner}. An analogy with X(5), which describes the critical point of a first order shape phase transition from spherical vibrator to $\gamma$-soft prolate nuclei, can be also made.

\begin{figure}
\begin{center}
\includegraphics[width=1\textwidth]{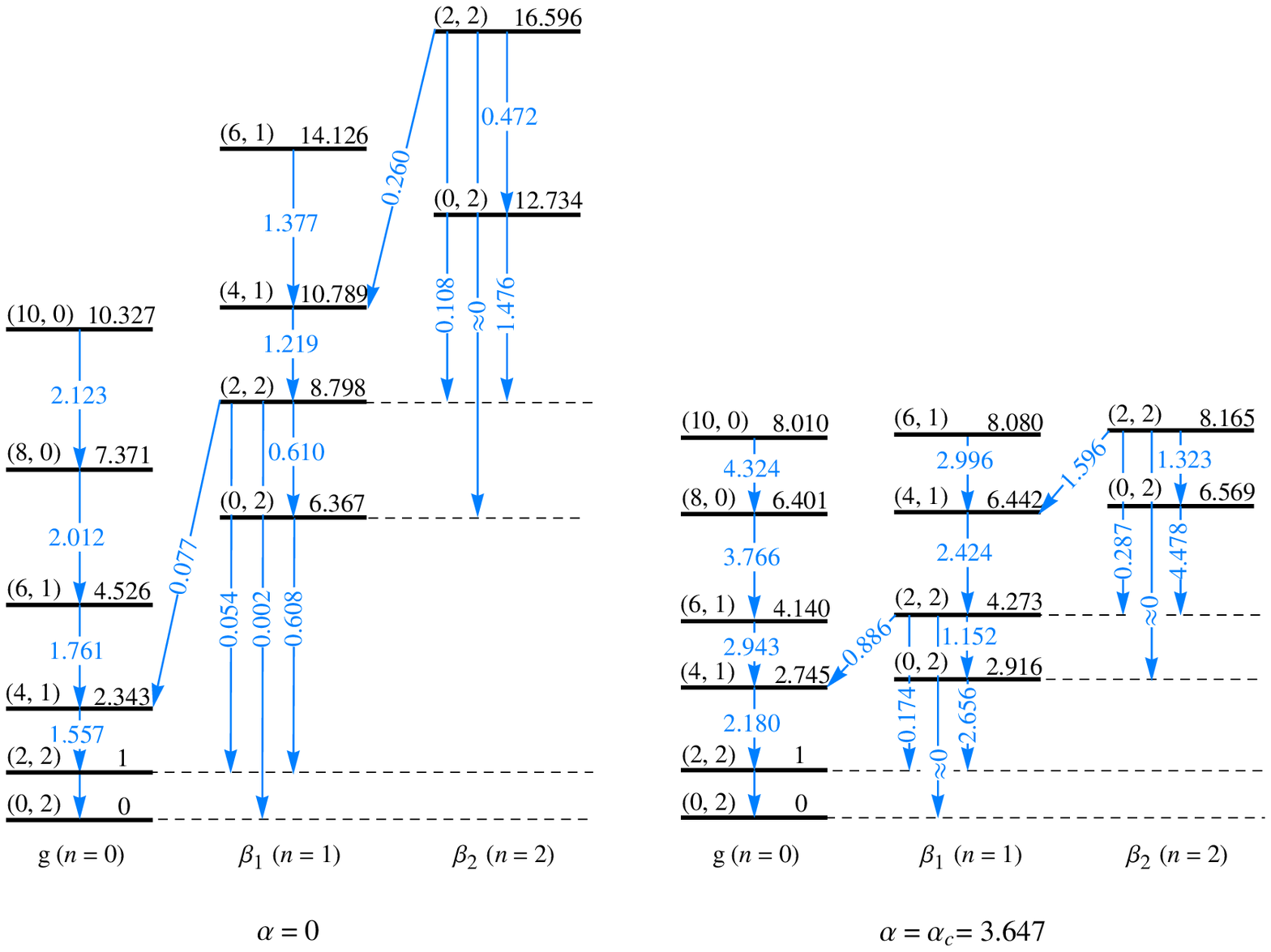}
\end{center}
\caption{Energy spectra (\ref{enrat}) and $B(E2)$ transitions (\ref{tranrap}) for $\alpha=0$ (left side) and $\alpha=\alpha_c$ (right side). The states are indexed by $(L,K)$.}
\label{fig3}
\end{figure}

By analyzing the expression of the sextic potential (\ref{poty})  it can be observed that the terms $y^{2}$ and $y^{4}$ are cancelling  for $\alpha=\pm2(c_{m}^{(K)})^{1/2}$ and $\alpha=0$, respectively, leading to free parameter solutions. The energy spectra (\ref{enrat}) and some $E2$ transitions in these particular cases
are presented in Fig. \ref{fig3} for $K=2$. If the energy spectrum for $\alpha=0$ doesn't show nothing  out of common, that for $\alpha_c$ is quite interesting from the point of view of its structure. Firstly, the states are grouped two by two in each band, and secondly an approximate degeneracy appears for states of different angular momenta ($\Delta L=4$) belonging to different bands.  Also the $B(E2)$ transitions are more intense for $\alpha=\alpha_c$ than $\alpha=0$, while the energies  are higher for $\alpha=0$ than those for $\alpha=\alpha_c$. An interpretation of the results at $\alpha_c$ in terms of some symmetries would be an interesting subject, especially that the dynamical symmetry of X(5) is still unknown. It should be noted that degeneracies in the critical regions have been also indicated in Refs. \cite{Casperson,Williams}, while an Euclidian dynamical symmetry unifying several of them has been proposed in Ref. \cite{Zhang}.

In order to verify the consistency of the approximation (\ref{sl02}), several energy ratios (\ref{enrat}) are compared in Table \ref{tab2} with those corresponding to X(3)-$\beta^{2}$ \cite{Budaca1}, X(3)-$\beta^{4}$ \cite{Budaca1}, X(3)-$\beta^{6}$ \cite{Budaca2} and X(3) \cite{Bonatsos3}, respectively. The resulted root mean square (rms) value is increasing as one goes to higher anharmonic solution.  As was expected, the results of X(3)-Sextic are closer to those of X(3)-$\beta^{2}$, X(3)-$\beta^{4}$, and X(3)-$\beta^{6}$ because it contains all these terms, while for a fidel reproduction of the ISWP the inclusion of higher anharmonicities is necessary  \cite{Bonatsos4}.

\setlength{\tabcolsep}{6.5pt}
\begin{table}
\caption{Several energy ratios (\ref{enrat}) given by X(3)-Sextic are compared with the predictions of the parameter free models X(3)-$\beta^{2}$ \cite{Budaca1}, X(3)-$\beta^4$ \cite{Budaca1}, X(3)-$\beta^{6}$ \cite{Budaca2} and X(3) \cite{Bonatsos3}.}
\vspace{0.3cm}
\centering
\def\arraystretch{0.3}
{\footnotesize
\begin{center}
\begin{tabular}{lcccccccccccc}
\hline
\hline\noalign{\smallskip}
Model&R$_{0,4}$&R$_{0,6}$&R$_{0,8}$&R$_{0,10}$&R$_{1,0}$&R$_{1,2}$&R$_{1,4}$&R$_{1,6}$&R$_{2,0}$&R$_{2,2}$&rms&$\alpha$\\
\noalign{\smallskip}\hline\noalign{\smallskip}
X(3)-$\beta^{2}$&2.13&3.27&4.42&5.58&2.00&3.00&4.13&5.27&4.00&5.00&&\\
                &2.11&3.24&4.40&5.56&2.02&3.02&4.11&5.21&4.05&5.06&0.04&25.6\\
\noalign{\medskip}
X(3)-$\beta^{4}$&2.29&3.72&5.28&6.94&2.37&3.70&5.29&6.98&5.28&6.83&&\\
                &2.45&3.72&5.49&6.89&2.48&3.67&5.36&6.76&5.40&6.73&0.13&4.92\\
\noalign{\medskip}
X(3)-$\beta^{6}$&2.34&3.91&5.65&7.57&2.56&4.08&5.94&7.97&6.01&7.93&&\\
                &2.65&3.99&6.08&7.61&2.76&4.05&6.05&7.59&6.14&7.63&0.24&4.04\\
\noalign{\smallskip}
X(3)&2.44&4.23&6.35&8.78&2.87&4.83&7.37&10.29&7.65&10.56&&\\
    &2.62&4.29&6.81&8.87&3.40&5.19&7.52&9.72&7.97&10.25&0.35&1.92\\
\noalign{\smallskip}
\hline
\hline
\end{tabular}\end{center}}
\label{tab2}
\end{table}

\subsection{Applications of X(3)-Sextic to experimental data}

The X(3)-Sextic solution is applied to describe the available experimental data for the ground and the first two $\beta$ bands of 39 nuclei, namely,  $^{98-108}$Ru, $^{100,102}$Mo, $^{116-130}$Xe, $^{132,134}$Ce, $^{146-150}$Nd, $^{150,152}$Sm, $^{152,154}$Gd, $^{154,156}$Dy, $^{172}$Os, $^{180-196}$Pt, $^{190}$Hg and $^{222}$Ra. The comparison for the energy spectra is given in Tables \ref{tab3} and \ref{tab4}, while for the $E2$ transitions is shown in Tables \ref{tab5} and \ref{tab6}. All these data are fitted involving only the free parameter $\alpha$, its values and the corresponding rms being indicated in the same tables. The agreement of X(3)-Sextic with the experimental data is good in general for most of the considered nuclei and in particular for $^{98-106}$Ru, $^{100}$Mo,  $^{122,128}$Xe, $^{134}$Ce, $^{148}$Nd, $^{150}$Sm, $^{188-196}$Pt and $^{190}$Hg for which small rms values are obtained. Instead, the experimental data of $^{108}$Ru, $^{116}$Xe, $^{132}$Ce, $^{150}$Nd, $^{152}$Sm, $^{154}$Gd, $^{156}$Dy, $^{172}$Os and $^{182}$Pt are well reproduced only partially, either the band heads or only the $\beta$ bands being well described. For the states where the experimental data are of uncertain band, a selection has been done based on the corresponding X(3)-Sextic results. This is mostly done for the band head state of the second $\beta$ band and then for states of the first $\beta$ band. These results can be very useful to clarify the nature of these states. Also, an experimental energy ratio of $2.76$ is attributed to the band head state of the first $\beta$ band of $^{104}$Ru, for which
X(3)-Sextic predicts the value $2.92$. As a matter of fact, the $^{104}$Ru nucleus is important here because its corresponding $\alpha=3.62$ value is very close to the critical point $\alpha_c=3.65$. As has been discussed in the previous subsection, this value of $\alpha$ leads to a parameter free description. In the same category are also included the nuclei $^{120,126}$Xe and $^{222}$Ra, the last one being close to the $\alpha=0$ case. Another interesting study carried out here has as starting point the fact that  X(3)-Sextic can simulate the $\gamma$-rigid version of X(5). Therefore the experimental data of the most representative candidate nuclei for X(5), namely, $^{150}$Nd, $^{152}$Sm, $^{154}$Gd and $^{156}$Dy are analyzed in the frame of the present solution. It is found that the ground band of these nuclei prefers a $\gamma$-stable structure, while the situation of the first two $\beta$ bands is changed in a clear favor for $\gamma$-rigidity. This behavior, also reported in connection to X(3) \cite{Bonatsos3},  reflects the fact that a combined scheme of $\gamma$-stable and $\gamma$-rigid  would be more appropriate for these nuclei. Actually, an important step toward this direction has already been done in Ref. \cite{Budaca3}, where a free parameter manages the $\gamma$-rigidity.

Concerning the $B(E2)$ transition rates, given in Tables \ref{tab5} and \ref{tab6}, the agreement of X(3)-Sextic with the corresponding experimental data is good taking into account the fact that the free parameter $\alpha$ was fitted only for the energy spectra. As can be seen, there are some discrepancies in the ground band of some nuclei for which X(3)-Sextic predicts increasing values in respect to $L$ while experimental data manifest a decreasing trend. This problem can be partly removed by including anharmonicities in the transition operator (\ref{tranop}) as in Ref. \cite{Raduta2}. With this, the $B(E2)$ results between states from different bands can also be improved.

\setlength{\tabcolsep}{6.7pt}
\begin{table}
\caption{The energy spectra for the ground band and the first two $\beta$ bands given by Eq. (\ref{enrat}) are compared with the available experimental data \cite{98Ru,100RuMo,102RuMo,104Ru,106Ru,108Ru,116Xe,118Xe,120Xe,122Xe,124Xe,126Xe,128Xe,130Xe,132Ce,134Ce} of the nuclei $^{98-108}$Ru, $^{100,102}$Mo, $^{116-130}$Xe and $^{132,134}$Ce. In the first line of each nucleus are given the experimental data, while in the second line are the corresponding theoretical results. In brackets are indicated possible candidate energy states for the corresponding predicted data, which were not included in the fit.}
\vspace{0.3cm}
\centering
\def\arraystretch{0.3}
{\footnotesize
\begin{tabular}{ccccccccccccc}
\hline
\hline\noalign{\smallskip}
Nucleus&R$_{0,4}$&R$_{0,6}$&R$_{0,8}$&R$_{0,10}$&R$_{1,0}$&R$_{1,2}$&R$_{1,4}$&R$_{1,6}$&R$_{2,0}$&R$_{2,2}$&rms&$\alpha$\\
\noalign{\smallskip}\hline\noalign{\smallskip}
$^{98}_{44}$Ru$_{54}$&2.14&3.41&4.79&&2.03&&&&&&&\\
          &2.21&3.38&4.73&5.97&2.15&3.21&4.49&5.68&4.46&5.57&0.08&8.44\\
\noalign{\medskip}
$^{100}_{44}$Ru$_{56}$&2.27&3.85&5.67&7.85&2.10&&&&&&&\\
          &2.58&3.90&5.88&7.36&2.66&3.92&5.82&7.31&5.89&7.32&0.37&4.28\\
\noalign{\medskip}
$^{102}_{44}$Ru$_{58}$&2.33&3.94&5.70&7.23&1.99&&&&&&&\\
          &2.50&3.79&5.64&7.08&2.55&3.77&5.54&6.98&5.59&6.96&0.28&4.63\\
\noalign{\medskip}
$^{104}_{44}$Ru$_{60}$&2.48&4.35&6.48&8.69&(2.76)&4.23&5.81&&&&&\\
          &2.74&4.14&6.41&8.02&2.92&4.28&6.45&8.10&6.58&8.19&0.40&3.62\\
\noalign{\medskip}
$^{106}_{44}$Ru$_{62}$&2.66&4.80&7.31&10.02&3.67&&&&&&&\\
          &2.55&4.43&7.14&9.55&4.10&6.20&8.55&11.19&9.36&12.20&0.34&1.16\\
\noalign{\medskip}
$^{108}_{44}$Ru$_{64}$&2.75&5.12&8.02&11.31&4.03&&&&&&&\\
          &2.49&4.50&7.30&9.92&4.75&7.01&9.31&12.22&10.43&13.65&0.83&0.73\\
\noalign{\medskip}
$^{100}_{42}$Mo$_{58}$&2.12&3.45&4.91&6.29&1.30&2.73&&&&&&\\
          &2.20&3.37&4.72&5.95&2.14&3.20&4.47&5.66&4.44&5.54&0.43&8.63\\
\noalign{\medskip}
$^{102}_{42}$Mo$_{60}$&2.51&4.48&6.81&9.41&2.35&3.86&&&&&&\\
          &2.67&4.19&6.57&8.39&3.09&4.65&6.90&8.80&7.16&9.07&0.63&2.65\\
\noalign{\medskip}
$^{116}_{54}$Xe$_{62}$&2.33&3.90&5.62&7.45&&2.58&3.96&5.38&&&&\\
          &2.28&3.48&4.97&6.26&2.25&3.35&4.76&6.02&4.75&5.93&0.72&6.62\\
\noalign{\medskip}
$^{118}_{54}$Xe$_{64}$&2.40&4.14&6.15&8.35&2.46&3.64&5.13&&(5.10)&&&\\
          &2.61&3.94&5.98&7.48&2.71&3.98&5.93&7.44&6.01&7.47&0.49&4.16\\
\noalign{\medskip}
$^{120}_{54}$Xe$_{66}$&2.47&4.33&6.51&8.90&2.82&3.95&5.31&&(6.93)&&&\\
          &2.73&4.11&6.32&7.91&2.88&4.21&6.34&7.94&6.45&8.01&0.57&3.79\\
\noalign{\medskip}
$^{122}_{54}$Xe$_{68}$&2.50&4.43&6.69&9.18&3.47&4.51&&&(7.63)&&&\\
          &2.63&4.26&6.74&8.73&3.29&5.02&7.33&9.44&7.72&9.89&0.30&2.11\\
\noalign{\medskip}
$^{124}_{54}$Xe$_{70}$&2.48&4.37&6.58&8.96&3.58&4.60&5.69&&(6.70)&&&\\
          &2.70&4.17&6.50&8.23&3.01&4.48&6.70&8.50&6.91&8.69&0.53&2.99\\
\noalign{\medskip}
$^{126}_{54}$Xe$_{72}$&2.42&4.21&6.27&8.64&3.38&4.32&5.25&&(6.57)&&&\\
          &2.72&4.09&6.29&7.86&2.86&4.18&6.30&7.89&6.40&7.95&0.55&3.83\\
\noalign{\medskip}
$^{128}_{54}$Xe$_{74}$&2.33&3.92&5.67&7.60&3.57&4.52&&&(5.87)&&&\\
          &2.66&4.01&6.12&7.65&2.78&4.07&6.10&7.65&6.19&7.69&0.44&4.00\\
\noalign{\medskip}
$^{130}_{54}$Xe$_{76}$&2.25&3.63&5.03&&3.35&(4.01)&(4.53)&&&&&\\
          &2.41&3.67&5.37&6.75&2.43&3.60&5.23&6.59&5.26&6.55&0.50&5.18\\
\noalign{\medskip}
$^{132}_{58}$Ce$_{74}$&2.64&4.74&7.16&9.71&3.56&4.60&5.94&&&&&\\
          &2.65&4.23&6.65&8.56&3.18&4.83&7.11&9.11&7.43&9.47&0.70&2.36\\
\noalign{\medskip}
$^{134}_{58}$Ce$_{76}$&2.56&4.55&6.87&9.09&3.75&4.80&&&&&&\\
          &2.61&4.30&6.82&8.89&3.42&5.23&7.56&9.77&8.01&10.31&0.26&1.89\\
\noalign{\medskip}
\hline
\hline
\end{tabular}}
\label{tab3}
\end{table}

\setlength{\tabcolsep}{6.7pt}
\begin{table}
\caption{The same as in Table \ref{tab3}, but for the available experimental data \cite{146Nd,148Nd,150NdSm,152SmGd,154GdDy,156Dy,172Os,180Pt,182Pt,184Pt,186Pt,188Pt,190PtHg,192Pt,194Pt,196Pt,222Ra} of the nuclei $^{146-150}$Nd, $^{150,152}$Sm, $^{152,154}$Gd, $^{154,156}$Dy, $^{172}$Os, $^{180-196}$Pt, $^{190}$Hg and $^{222}$Ra.}
\vspace{0.3cm}
\centering
\def\arraystretch{0.3}
{\footnotesize
\begin{tabular}{ccccccccccccc}
\hline
\hline\noalign{\smallskip}
Nucleus&R$_{0,4}$&R$_{0,6}$&R$_{0,8}$&R$_{0,10}$&R$_{1,0}$&R$_{1,2}$&R$_{1,4}$&R$_{1,6}$&R$_{2,0}$&R$_{2,2}$&rms&$\alpha$\\
\noalign{\smallskip}\hline\noalign{\smallskip}
$^{146}_{60}$Nd$_{86}$&2.30&3.92&5.72&7.32&2.02&2.87&3.85&&&&&\\
          &2.34&3.56&5.14&6.46&2.32&3.45&4.96&6.26&4.96&6.19&0.64&5.88\\
\noalign{\medskip}
$^{148}_{60}$Nd$_{88}$&2.49&4.24&6.15&8.19&3.04&3.88&5.32&7.12&(5.30)&&&\\
          &2.61&3.95&5.98&7.49&2.71&3.98&5.94&7.46&6.02&7.48&0.39&4.15\\
\noalign{\medskip}
$^{150}_{60}$Nd$_{90}$&2.93&5.53&8.68&12.28&5.19&6.53&8.74&11.83&(13.35)&&&\\
          &2.49&4.50&7.30&9.92&4.75&7.01&9.31&12.22&10.43&13.65&1.10&0.73\\
\noalign{\medskip}
$^{150}_{62}$Sm$_{88}$&2.32&3.83&5.50&7.29&2.22&3.13&4.34&6.31&(3.76)&&&\\
          &2.36&3.60&5.23&6.57&2.36&3.51&5.06&6.39&5.07&6.33&0.41&5.58\\
\noalign{\medskip}
$^{152}_{62}$Sm$_{90}$&3.01&5.80&9.24&13.21&5.62&6.65&8.40&10.76&8.89&10.62&&\\
          &2.55&4.43&7.14&9.55&4.10&6.20&8.55&11.19&9.36&12.20&1.59&1.16\\
\noalign{\medskip}
$^{152}_{64}$Gd$_{88}$&2.19&3.57&5.07&6.68&1.79&2.70&3.72&4.85&3.04&3.83&&\\
          &2.10&3.22&4.37&5.52&2.00&3.00&4.06&5.16&4.00&5.00&0.68&100\\
\noalign{\medskip}
$^{154}_{64}$Gd$_{90}$&3.01&5.83&9.30&13.3&5.53&6.63&8.51&11.10&9.60&11.52&&\\
          &2.53&4.46&7.21&9.70&4.33&6.50&8.84&11.58&9.75&12.74&1.51&0.99\\
\noalign{\medskip}
$^{154}_{66}$Dy$_{88}$&2.23&3.66&5.23&6.89&1.98&2.71&3.74&4.96&3.16&4.16&&\\
          &2.10&3.22&4.37&5.52&2.00&3.00&4.06&5.16&4.00&5.00&0.67&100\\
\noalign{\medskip}
$^{156}_{66}$Dy$_{90}$&2.93&5.59&8.82&12.52&4.90&6.01&7.90&10.43&10.00&&&\\
          &2.53&4.46&7.20&9.68&4.31&6.46&8.81&11.53&9.71&12.68&1.28&1.01\\
\noalign{\smallskip}
$^{172}_{76}$Os$_{96}$&2.66&4.63&6.70&8.89&3.33&3.56&5.00&6.81&&&&\\
          &2.65&4.00&6.10&7.63&2.77&4.06&6.07&7.62&6.16&7.66&0.77&4.02\\
\noalign{\medskip}
$^{180}_{78}$Pt$_{102}$&2.68&4.94&7.71&10.93&3.12&5.62&8.15&10.77&(7.69)&&&\\
          &2.56&4.40&7.07&9.41&3.92&5.96&8.32&10.86&9.03&11.75&0.69&1.31\\
\noalign{\medskip}
$^{182}_{78}$Pt$_{104}$&2.71&5.00&7.78&10.96&3.22&5.53&8.00&10.64&(7.43)&&&\\
          &2.57&4.40&7.05&9.37&3.88&5.90&8.26&10.77&8.95&11.64&0.72&1.35\\
\noalign{\medskip}
$^{184}_{78}$Pt$_{106}$&2.67&4.90&7.55&10.47&3.02&5.18&7.57&11.04&&&&\\
          &2.57&4.38&7.01&9.29&3.79&5.77&8.13&10.59&8.77&11.39&0.65&1.44\\
\noalign{\medskip}
$^{186}_{78}$Pt$_{108}$&2.56&4.58&7.01&9.70&2.46&4.17&6.38&8.36&&&&\\
          &2.68&4.18&6.53&8.30&3.04&4.56&6.80&8.64&7.03&8.87&0.62&2.82\\
\noalign{\medskip}
$^{188}_{78}$Pt$_{110}$&2.53&4.46&6.71&9.18&3.01&4.20&&&&&&\\
          &2.65&4.22&6.64&8.53&3.17&4.80&7.08&9.06&7.39&9.41&0.38&2.40\\
\noalign{\medskip}
$^{190}_{78}$Pt$_{112}$&2.49&4.35&6.47&8.57&3.11&4.07&&&(5.65)&&&\\
          &2.69&4.17&6.50&8.24&3.01&4.50&6.72&8.52&6.93&8.72&0.25&2.96\\
\noalign{\medskip}
$^{192}_{78}$Pt$_{114}$&2.48&4.31&6.38&8.62&3.78&4.55&&&&&&\\
          &2.65&4.23&6.66&8.57&3.19&4.84&7.13&9.14&7.45&9.50&0.30&2.34\\
\noalign{\medskip}
$^{194}_{78}$Pt$_{116}$&2.47&4.30&6.39&8.67&3.23&4.60&&&&&&\\
          &2.64&4.24&6.69&8.63&3.86&4.91&7.20&9.25&7.55&9.65&0.32&2.25\\
\noalign{\medskip}
$^{196}_{78}$Pt$_{118}$&2.47&4.29&6.33&8.56&3.19&3.83&&&&&&\\
          &2.72&4.15&6.45&8.13&2.96&4.39&6.59&8.31&6.76&8.46&0.33&3.25\\
\noalign{\medskip}
$^{190}_{80}$Hg$_{110}$&2.50&4.26&&&3.07&3.77&4.74&6.03&&&&\\
          &2.36&3.59&5.21&6.56&2.36&3.50&5.05&6.37&5.06&6.31&0.46&5.62\\
\noalign{\medskip}
$^{222}_{88}$Ra$_{134}$&2.71&4.95&7.59&10.56&8.23&9.22&&&&&&\\
          &2.25&4.48&7.29&10.37&7.44&9.90&11.63&15.14&14.19&18.33&0.52&-0.38\\
\noalign{\smallskip}
\hline
\hline
\end{tabular}}
\label{tab4}
\end{table}

\setlength{\tabcolsep}{6.5pt}
\begin{table}[th!]
\caption{Several $B(E2)$ transition rates normalized as in Eq. (\ref{tranrap}) are compared with the available experimental data \cite{98Ru,100RuMo,102RuMo,104Ru,106Ru,108Ru,116Xe,118Xe,120Xe,122Xe,124Xe,126Xe,128Xe,130Xe,132Ce,134Ce} of the nuclei $^{98-108}$Ru, $^{100,102}$Mo, $^{116-130}$Xe, $^{132,134}$Ce. In the first line of each nucleus are given the experimental data, while in each second line are the corresponding theoretical results.}
\vspace{0.3cm}
\begin{center}
\def\arraystretch{0.3}
{\footnotesize
\begin{tabular}{cllllllll}
\hline
\hline\noalign{\smallskip}
Nucleus&$T_{0,4,0,2}$& $T_{0,6,0,4}$& $T_{0,8,0,6}$& $T_{0,10,0,8}$& $T_{1,0,0,2}$& $T_{1,2,0,2}$& $T_{1,2,0,4}$& $T_{1,2,1,0}$\\
\noalign{\smallskip}\hline\noalign{\smallskip}
$^{98}_{44}$Ru$_{54}$ &0.38(11)&0.40(8)&0.08(2)&&&&&\\
          &2.34&3.48&4.62&5.62&3.21&0.21&1.47&1.52\\
\noalign{\medskip}
$^{100}_{44}$Ru$_{56}$&1.43(11)&4.78(5)&&&0.98(14)&&&\\
          &2.23&3.08&3.98&4.63&2.83&0.18&1.02&1.24\\
\noalign{\medskip}
$^{102}_{44}$Ru$_{58}$&1.48(25)&1.52(56)&1.26(43)&1.28(47)&0.78(14)&&&\\
          &2.25&3.15&4.08&4.77&2.90&0.19&1.09&1.28\\
\noalign{\medskip}
$^{104}_{44}$Ru$_{60}$&1.43(16)&&&0.43(5)&&&&\\
          &2.18&2.94&3.76&4.31&2.65&0.17&0.88&1.15\\
\noalign{\medskip}
$^{106}_{44}$Ru$_{62}$&&&&&&&&\\
          &1.75&2.07&2.47&2.67&1.22&0.09&0.24&0.69\\
\noalign{\medskip}
$^{108}_{44}$Ru$_{64}$&1.65(20)&&&&&&&\\
          &1.66&1.93&2.27&2.43&0.95&0.08&0.16&0.64\\
\noalign{\medskip}
$^{100}_{42}$Mo$_{58}$&1.86(11)&2.54(38)&3.32(49)&&2.49(12)&$\approx0$&0.97(49)&0.38(11)\\
          &2.35&3.49&4.64&5.64&3.21&0.21&1.48&1.53\\
\noalign{\medskip}
$^{102}_{42}$Mo$_{60}$&1.20(28)&&&0.95(42)&&&&\\
          &2.05&2.64&3.31&3.71&2.23&0.15&0.62&0.97\\
\noalign{\medskip}
$^{116}_{54}$Xe$_{62}$&1.75(11)&1.58(15)&&1.56(29)&&&&\\
          &2.32&3.38&4.45&5.34&3.12&0.20&1.35&1.44\\
\noalign{\medskip}
$^{118}_{54}$Xe$_{64}$&1.11(6)&0.88(23)&0.49(18)&$>$0.7&&&&\\
          &2.22&3.06&3.95&4.58&2.80&0.18&1.00&1.23\\
\noalign{\medskip}
$^{120}_{54}$Xe$_{66}$&1.16(10)&1.17(19)&0.96(17)&0.91(16)&&&&\\
          &2.19&2.98&3.82&4.40&2.70&0.18&0.91&1.17\\
\noalign{\medskip}
$^{122}_{54}$Xe$_{68}$&1.46(11)&1.41(9)&1.03(8)&1.54(10)&&&&\\
          &1.95&2.44&3.01&3.33&1.90&0.13&0.47&0.86\\
\noalign{\medskip}
$^{124}_{54}$Xe$_{70}$&1.17(4)&1.52(14)&1.14(36)&0.36(5)&&&&\\
          &2.10&2.76&3.48&3.94&2.40&0.16&0.72&1.04\\
\noalign{\medskip}
$^{126}_{54}$Xe$_{72}$&&&&&&&&\\
          &2.20&2.99&3.83&4.42&2.71&0.18&0.93&1.18\\
\noalign{\medskip}
$^{128}_{54}$Xe$_{74}$&1.47(15)&1.94(20)&2.39(30)&&&&&\\
          &2.21&3.03&3.89&4.50&2.76&0.18&0.97&1.20\\
\noalign{\medskip}
$^{130}_{54}$Xe$_{76}$&&&&&&&&\\
          &2.28&3.23&4.22&4.97&2.99&0.19&1.18&1.34\\
\noalign{\medskip}
$^{132}_{58}$Ce$_{74}$&1.11(26)&1.51(87)&0.73(16)&0.47(12)&&&&\\
          &2.00&2.53&3.15&3.51&2.06&0.14&0.54&0.91\\
\noalign{\medskip}
$^{134}_{58}$Ce$_{76}$&0.75(17)&0.27(10)&0.47(4)&0.07(2)&&&&\\
          &1.91&2.35&2.88&3.17&1.75&0.12&0.41&0.82\\
\noalign{\medskip}
\hline
\hline
\end{tabular}}
\end{center}
\label{tab5}
\end{table}

\setlength{\tabcolsep}{6.5pt}
\begin{table}[th!]
\caption{The same as in Table \ref{tab5}, but for the available experimental data \cite{146Nd,148Nd,150NdSm,152SmGd,154GdDy,156Dy,172Os,180Pt,182Pt,184Pt,186Pt,188Pt,190PtHg,192Pt,194Pt,196Pt,222Ra} of the nuclei $^{146-150}$Nd, $^{150,152}$Sm, $^{152,154}$Gd, $^{154,156}$Dy, $^{172}$Os, $^{180-196}$Pt, $^{190}$Hg and $^{222}$Ra.
}
\vspace{0.3cm}
\begin{center}
\def\arraystretch{0.3}
{\footnotesize
\begin{tabular}{cllllllll}
\hline
\hline\noalign{\smallskip}
Nucleus&$T_{0,4,0,2}$& $T_{0,6,0,4}$& $T_{0,8,0,6}$& $T_{0,10,0,8}$& $T_{1,0,0,2}$& $T_{1,2,0,2}$& $T_{1,2,0,4}$& $T_{1,2,1,0}$\\
\noalign{\smallskip}\hline\noalign{\smallskip}
$^{146}_{60}$Nd$_{86}$&1.47(39)&&&&&&&\\
          &2.30&3.31&4.35&5.17&3.06&0.20&1.27&1.40\\
\noalign{\medskip}
$^{148}_{60}$Nd$_{88}$&1.62(9)&1.76(14)&1.69(30)&&0.54(4)&0.25(3)&0.28(14)&\\
          &2.22&3.06&3.94&4.57&2.80&0.18&1.00&1.23\\
\noalign{\medskip}
$^{150}_{60}$Nd$_{90}$&1.56(4)&1.78(9)&1.86(20)&1.73(10)&0.37(2)&0.09(3)&0.16(6)&1.38(112)\\
          &1.66&1.93&2.27&2.43&0.95&0.08&0.16&0.64\\
\noalign{\medskip}
$^{150}_{62}$Sm$_{88}$&1.93(30)&2.63(88)&2.98(158)&&0.93(9)&&&1.93$^{+0.70}_{-0.53}$\\
          &2.29&3.28&4.30&5.09&3.03&0.20&1.23&1.37\\
\noalign{\medskip}
$^{152}_{62}$Sm$_{90}$&1.44(2)&1.66(3)&2.02(4)&2.17$^{+0.24}_{-0.18}$&0.23(1)&0.04&0.12(1)&1.17(8)\\
          &1.75&2.07&2.47&2.67&1.22&0.09&0.24&0.69\\
\noalign{\medskip}
$^{152}_{64}$Gd$_{88}$&1.82$^{+0.20}_{-0.18}$&2.70$^{+0.72}_{-0.53}$&2.44$^{+0.76}_{-0.49}$&&&0.23(3)&0.32(7)&0.49(7)\\
          &2.38&3.67&4.94&6.20&3.33&0.22&1.71&1.67\\
\noalign{\medskip}
$^{154}_{64}$Gd$_{90}$&1.56(6)&1.82(10)&1.99(11)&2.29(26)&0.33(5)&0.04&0.12(1)&0.62(6)\\
          &1.71&2.01&2.39&2.57&1.11&0.09&0.20&0.67\\
\noalign{\medskip}
$^{154}_{66}$Dy$_{88}$&1.43(15)&1.98(44)&1.98(44)&1.98(64)&&&&\\
          &2.38&3.67&4.94&6.20&3.33&0.22&1.71&1.67\\
\noalign{\medskip}
$^{156}_{66}$Dy$_{90}$&1.63(2)&1.87(6)&2.07(20)&2.20(27)&&0.09(3)&0.08(3)&\\
          &1.71&2.02&2.40&2.58&1.12&0.09&0.21&0.67\\
\noalign{\smallskip}
$^{172}_{76}$Os$_{96}$&1.50(17)&2.61(38)&3.30(98)&1.65(36)&&&&\\
          &2.21&3.03&3.90&4.51&2.76&0.18&0.97&1.21\\
\noalign{\medskip}
$^{180}_{78}$Pt$_{102}$&0.92(22)&$\geq$0.29&&&&&&\\
          &1.78&2.12&2.55&2.77&1.33&0.10&0.27&0.72\\
\noalign{\medskip}
$^{182}_{78}$Pt$_{104}$&&&&&&&&\\
          &1.79&2.14&2.57&2.80&1.36&0.10&0.28&0.72\\
\noalign{\medskip}
$^{184}_{78}$Pt$_{106}$&1.65(9)&1.78(12)&2.13(16)&2.44(33)&&&&\\
          &1.81&2.17&2.62&2.86&1.42&0.10&0.30&0.74\\
\noalign{\medskip}
$^{186}_{78}$Pt$_{108}$&&&&&&&&\\
          &2.08&2.70&3.40&3.83&2.32&0.15&0.67&1.01\\
\noalign{\medskip}
$^{188}_{78}$Pt$_{110}$&&&&&&&&\\
          &2.01&2.55&3.17&3.54&2.08&0.14&0.55&0.92\\
\noalign{\medskip}
$^{190}_{78}$Pt$_{112}$&&&&&&&&\\
          &2.10&2.75&3.46&3.92&2.39&0.16&0.71&1.03\\
\noalign{\medskip}
$^{192}_{78}$Pt$_{114}$&1.56(9)&1.22(53)&&&&&&\\
          &2.00&2.53&3.14&3.50&2.05&0.14&0.54&0.91\\
\noalign{\medskip}
$^{194}_{78}$Pt$_{116}$&1.73(11)&1.36(43)&1.02(29)&0.69(18)&&&&\\
          &1.98&2.49&3.09&3.43&1.99&0.14&0.51&0.89\\
\noalign{\medskip}
$^{196}_{78}$Pt$_{118}$&1.48(2)&1.80(10)&1.92(25)&&&&$\approx0$&0.12(12)\\
          &2.14&2.83&3.60&4.10&2.51&0.17&0.79&1.09\\
\noalign{\medskip}
$^{190}_{80}$Hg$_{110}$&&&&&&&&\\
          &2.29&3.28&4.30&5.10&3.04&0.20&1.24&1.38\\
\noalign{\medskip}
$^{222}_{88}$Ra$_{134}$&&&&&&&&\\
          &1.52&1.71&1.92&2.02&0.49&0.05&0.05&0.62\\
\noalign{\smallskip}
\hline
\hline
\end{tabular}}
\end{center}
\label{tab6}
\end{table}

\begin{figure}
\begin{center}
\includegraphics[width=1\textwidth]{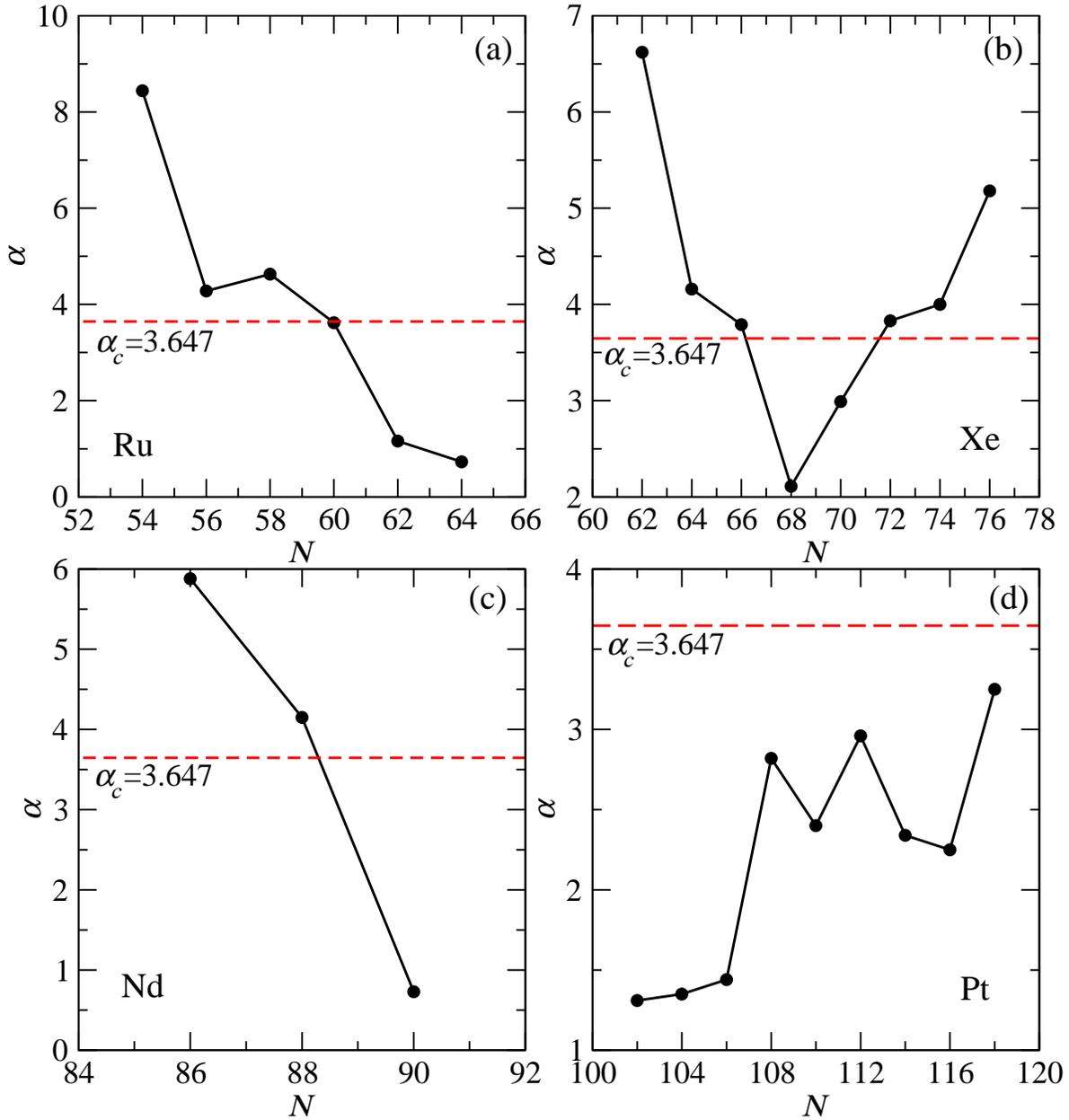}
\end{center}
\caption{The plot of the free parameter $\alpha$ as a function of the neutron number $N$ for the isotopes of Ru (a), Xe (b), Nd (c) and Pt (d).}
\label{fig4}
\end{figure}

Another important aspect that can be investigated by X(3)-Sextic is whether or not a shape phase transition takes place within an isotopic chain.  For this purpose, the fitted values of $\alpha$  are plotted in Fig. \ref{fig4} as a function of the neutron number $N$ for the most numerous isotopic chains considered in the present paper, namely Ru, Xe, Nd and Pt.
In Fig. \ref{fig4}, the regions above and bellow the dashed line indicating the critical point are equivalent in Fig. \ref{fig2} with  the right side and the left side, respectively, of the critical region represented by the gray narrow area.
The isotopes of Ru and Nd have a similar behavior crossing once the dashed line indicating the critical point and having the lightest isotopes and the heaviest ones situated above and below the line, respectively. By contrary, the isotopes of Xe cross twice the critical point, once from the lightest isotopes towards the medium ones ($^{116}_{54}$Xe$_{62}\rightarrow \,^{122}_{54}$Xe$_{68}$) and second time from the heaviest isotopes towards the medium ones ($^{130}_{54}$Xe$_{76}\rightarrow\, ^{122}_{54}$Xe$_{68}$). If the points $(\alpha,N)$ of the Xe isotopes are interpolated by a polynomial function, then the first derivative of the resulted function $\alpha=f(N)$ would present a discontinuity at the point of $^{122}_{54}$Xe$_{68}$ where the two separated shape phase transitions converge. This last observation can be interpreted as follows. If $^{120}_{54}$Xe$_{66}$ and $^{126}_{54}$Xe$_{72}$ are candidates for the critical points of the two shape phase transitions ($^{116}_{54}$Xe$_{62}\rightarrow\, ^{122}_{54}$Xe$_{68}$) and ($^{130}_{54}$Xe$_{76}\rightarrow\,^{122}_{54}$Xe$_{68}$), respectively, the isotope $^{122}_{54}$Xe$_{68}$ plays the role of a critical point of a transition between the two transition arms of the Xe isotopes. In other words the transition from $^{120}_{54}$Xe$_{66}$ to $^{124}_{54}$Xe$_{70}$ or $^{126}_{54}$Xe$_{72}$ is not smooth at all but rather is like a jump. This is not the case for the Ru isotopes or for those of Nd where the transition is smooth and unidirectional. Instead, a similar jump is observed for the isotopes of Pt, namely between $^{184}_{78}$Pt$_{106}$ and $^{186}_{78}$Pt$_{108}$. Nevertheless, because no crossing of the critical point takes place for these isotopes, the jump is associated  to a critical point of a local shape phase transition taking place between the group of more deformed isotopes $^{180-184}_{78}$Pt$_{102-106}$ and the less deformed group of $^{186-196}_{78}$Pt$_{108-118}$ nuclei.
Concerning the best candidates for the critical point of the $\gamma$-rigid prolate harmonic vibrator to $\gamma$-rigid anharmonic vibrator shape phase transition, by far it is the $^{104}_{44}$Ru$_{60}$ nucleus followed closely by $^{120,126}_{54}$Xe$_{66,72}$ and $^{148}_{60}$Nd$_{88}$. Surprisingly, $^{104}_{44}$Ru$_{60}$ was found to be a good candidate also for the critical point of the spherical to $\gamma$-unstable shape phase transition \cite{Levai2,Frank} and for a situation intermediate between the triaxial and the $\gamma$-unstable limit \cite{Bijker,Dieperink,Wenes }, but what is interesting in the frame of the present description is that the isotope $^{104}_{44}$Ru$_{60}$ falls exactly in the critical point of the new transition. Other good candidates for the critical point may be considered $^{128}_{54}$Xe$_{74}$ and even $^{196}_{78}$Pt$_{118}$ and $^{172}_{76}$Os$_{96}$ despite the fact that no crossing of the critical point have been observed within their chains. $^{128}_{54}$Xe$_{74}$ was also proposed in Ref. \cite{Clark} as a good candidate for the E(5) symmetry,  while in Refs. \cite{Kharb,Levai3} where a sextic potential is used to describe the shape phase transition from spherical vibrator to $\gamma$-unstable nuclei, $^{124}_{54}$Xe$_{70}$ and $^{122}_{54}$Xe$_{68}$ are indicated as  being the most closest to the corresponding critical point. All the above examples suggest that the shape phase transition predicted by X(3)-Sextic has a strong experimental support.

Before going to conclusions, some final remarks are worth to be made here in order to frame the present solution between other Bohr-Hamiltonian approaches.
X(3)-Sextic is adding to the special class of rigid solutions, which implies that one or more variables are frozen. Most of these solutions have the advantage of being exactly determined, obtained in an analytical form, not depending on free parameters and giving good agreement with the corresponding experimental data. Of course there are also some drawbacks, the major one being that by freezing some degrees of freedom a part of information is definitively lost. A good example in that sense is the impossibility to describe the $\gamma$ band. An exception is the Z(4) type solutions \cite{Bonatsos2,Buganu1}, where the description of $\gamma$ band is still possible. Nevertheless, this type of solutions should not be seen as stand alone descriptions but rather as complementary approaches. As has been discussed above, these solutions can be useful in the understanding of the critical point phenomena, in the clarifying of the nature of some ambiguous states and of course in the generating of new development perspectives for the present field or other branches of physics, respectively.

\section{Conclusions}

$\;\;\;\;\;\;\;\;$A new solution of the Bohr-Mottelson Hamiltonian \cite{Bohr1,Bohr2} with a sextic potential \cite{Ushveridze} is proposed, this time for $\gamma$-rigid prolate nuclei. The model is called X(3)-Sextic. The separation of variables is exactly achieved. The angular equation corresponds to the spherical harmonic functions, while that for the $\beta$ variable with  a sextic potential is reduced to a quasi-exactly solvable form \cite{Ushveridze}. Finally, the energy spectrum and the wave function are given in analytical form and moreover, up to a scale factor, depend on a single free parameter. In particular cases of the sextic potential, when $\beta^{2}$ or $\beta^{4}$ cancels, parameter free solutions are obtained. As the properties of the states $0^{+}$ and $2^{+}$ are exactly determined, the model becomes useful in deciding whether or not some ambiguous states are the heads of the first two $\beta$ bands.
The comparison of the X(3)-Sextic results with those of previous solutions as X(3)-$\beta^{2}$ \cite{Budaca1}, X(3)-$\beta^{4}$ \cite{Budaca2}, X(3)-$\beta^{6}$ \cite{Budaca2} and X(3) \cite{Bonatsos3} reveals a good agreement between them, establishing in this way that the approximations involved to describe states characterized by $L\geq4$ are rigorous enough.

Numerical applications are done for 39 nuclei presenting experimental data at least for the ground band and first $\beta$ band, namely for $^{98-108}$Ru, $^{100,102}$Mo, $^{116-130}$Xe, $^{132,134}$Ce, $^{146-150}$Nd, $^{150,152}$Sm, $^{152,154}$Gd, $^{154,156}$Dy, $^{172}$Os, $^{180-196}$Pt, $^{190}$Hg and $^{222}$Ra. For those nuclei where the band head states of the first two $\beta$ bands were uncertain, a selection has been done by choosing those of them which were closer to the predicted ones. From all these applications is evidenced not only the accuracy of X(3)-Sextic to reproduce the existing experimental data, but also its power of prediction.

X(3)-Sextic describes very well only the $\beta$ bands of some X(5) candidate nuclei, as $^{150}$Nd, $^{152}$Sm, $^{154}$Gd and $^{156}$Dy, acting as a complement of X(5) which describes very well the ground state band, the $\gamma$ bands and the head of the first $\beta$ band. Therefore, a combination of the two solutions as in Ref. \cite{Budaca3} by involving a rigidity control parameter seems to be a proper method for these nuclei.

By studying the dependence of the energies on the free parameter, a first order shape phase transition is pointed out covering a region between $\gamma$-rigid prolate harmonic vibrator and $\gamma$-rigid prolate anharmonic vibrator. This behavior is also experimentally evidenced within the isotopic chains of Ru, Xe and Nd, where the best candidates for the critical point are $^{104}$Ru, $^{120,126}$Xe and $^{148}$Nd, respectively. Other possible candidates can be also considered $^{196}$Pt, $^{172}$Os or $^{128}$Xe.

\section*{Appendix}

The eigenfunctions $P^{(M)}_{n,L}(y^{2})$ and the eigenvalues $\lambda_{n,L}^{(M)}$ of the quasi-exactly solvable equation (\ref{eqQ}) are given here in an explicit form for $M=0,1,2$.

The case $M=0$:
\begin{equation}
P_{0,L}^{(0)}(y^{2})=N_{0,L}^{(0)},\;\;\lambda_{0,L}^{(0)}=0.
\end{equation}

The case $M=1$:
\begin{eqnarray}
P_{n,L}^{(1)}(y^{2})&=&N_{n,L}^{(1)}\left[1-\frac{\lambda_{n,L}^{(1)}}{2(L+3)}y^{2}\right],\;n\in\{0,1\},\nonumber\\
\lambda_{0,L}^{(1)}&=&2\left[\alpha-\sqrt{\alpha^{2}+2(L+3)}\right],\\
\lambda_{1,L}^{(1)}&=&2\left[\alpha+\sqrt{\alpha^{2}+2(L+3)}\right].\nonumber
\end{eqnarray}

The case $M=2$:
\begin{eqnarray}
P_{n,L}^{(2)}(y^{2})&=&N_{n,L}^{(2)}\left[1-\frac{\lambda_{n,L}^{(2)}}{2(L+3)}y^{2}-\frac{2\lambda_{n,L}^{(2)}}{(L+3)(\lambda_{n,L}^{(2)}-8\alpha)}y^{4}\right],\;n\in\{0,1,2\},\nonumber\\
\lambda_{0,L}^{(2)}&=&4\alpha-\frac{2i(-i+\sqrt{3})(8+\alpha^{2}+2L)}{3^{\frac{1}{3}}D(L,\alpha)}+\frac{2i(i+\sqrt{3})D(L,\alpha)}{3^{\frac{2}{3}}},\nonumber\\
\lambda_{1,L}^{(2)}&=&4\alpha+\frac{2i(i+\sqrt{3})(8+\alpha^{2}+2L)}{3^{\frac{1}{3}}D(L,\alpha}-\frac{2i(-i+\sqrt{3})D(L,\alpha)}{3^{\frac{2}{3}}},\\
\lambda_{2,L}^{(2)}&=&4\alpha+\frac{4(8+\alpha^{2}+2L)}{3^{\frac{1}{3}}D(L,\alpha)}+\frac{4D(L,\alpha)}{3^{\frac{2}{3}}},\nonumber\\
D(L,\alpha)&=&\left[9\alpha+\sqrt{3}\sqrt{27\alpha^{2}-(8+\alpha^{2}+2L)^{3}}\right]^{\frac{1}{3}},\;\;i=\sqrt{-1}.\nonumber
\end{eqnarray}

Here $N_{n,L}^{(M)}$ are constants which are determined from the conditions that the ansatz functions (\ref{ansatz}) to be normalized to unity with the integration measure $dy$. These constants can be determined either numerically or analytically by evaluating integrals of the form \cite{Levai1}:
\begin{equation}
I_{A}=\int_{0}^{\infty}y^{A}e^{-\frac{y^{4}}{2}-\alpha y^{2}}dy=\frac{1}{2}\Gamma\left(\frac{A+1}{2}\right)2^{\frac{1-A}{4}}U\left(\frac{A+1}{4},\frac{1}{2};\frac{\alpha^{2}}{2}\right),
\end{equation}
where $U$ is a confluent hypergeometric function \cite{Abramowitz}. It is necessary to point out that all solutions are real defined due to the special relations realized between the parameters of the quasi-exactly solvable sextic potential.

\ack
The authors acknowledges the financial support received from the Romanian Ministry of Education and Research, through the Project PN-09-37-01-02.

\end{document}